\documentclass{osa-article}
\usepackage{booktabs}
\journal{oe} 

\articletype{Research Article}

\begin{document}

\title{Angular displacements estimation enhanced by squeezing and parametric amplification}

\author{Gao-Feng Jiao,\authormark{1,4} Qiang Wang,\authormark{1,4} L. Q. Chen,\authormark{1,5} Weiping Zhang,\authormark{2,3} and Chun-Hua Yuan\authormark{1,3,*}}

\address{\authormark{1}State Key Laboratory of Precision Spectroscopy, Quantum Institute for Light and Atoms, Department of Physics, East China Normal University, Shanghai 200062, China\\
\authormark{2}School of Physics and Astronomy, and Tsung-Dao Lee Institute, Shanghai Jiao Tong University, Shanghai 200240, China\\
\authormark{3}Collaborative Innovation Center of Extreme Optics, Shanxi University, Taiyuan, Shanxi 030006, China\\
\authormark{4}These two authors contributed equally to this work.\\
\authormark{5}lqchen@phy.ecnu.edu.cn}

\email{\authormark{*}chyuan@phy.ecnu.edu.cn} 



\begin{abstract}
We theoretically study the angular displacements estimation based on a modified
Mach-Zehnder interferometer (MZI), in which two optical parametric amplifiers (PAs)
are introduced into two arms of the standard MZI, respectively. The employment
of PAs can both squeeze the shot noise and amplify
the photon number inside the interferometer. When the unknown angular
displacements are introduced to both arms, we derive the multiparameter quantum Cram\'{e}r-Rao bound (QCRB) using the quantum Fisher information matrix approach, and the bound of angular displacements difference between the two arms is compared with the sensitivity of angular displacement using the intensity detection. On the other hand, in the case where the unknown angular displacement is in only one arm, we give the sensitivity of angular displacement using the method of homodyne
detection. It can surpass the standard quantum limit (SQL) and approach the
single parameter QCRB. Finally, the effect of photon losses on sensitivity is discussed.
\end{abstract}


\section{Introduction}

Phase estimation illustrates well the advantages of quantum metrology, with a
wide range of applications\cite{1,2,3,4,5}. As fundamental devices, a number
of interferometer configurations have been proposed for phase estimation. One
common configuration is the SU(2) interferometer, e.g., a Mach-Zehnder
interferometer (MZI), consists of two linear beam splitters (BSs). The
sensitivity of these interferometers is limited by the vacuum fluctuation
entering from the unused input port. To further enhance measurement
sensitivity, another commonly used configuration is proposed by Yurke
\textit{et al.}\cite{6}, known as the SU(1,1) interferometer, in which
optical parametric amplifiers (PAs) or four-wave mixers are employed as the wave
splitting and recombination elements. Because the signal is amplified while
the noise level is kept close to the shot noise limit, these types of
interferometers have been extensively studied both in
theory\cite{7,8,9,10,11,12,13,14,15,16,17} and
experiment\cite{18,19,20,21,22,23,24,25,26}. Besides the above, the
interferometers with novel structures have emerged in large numbers. Kong
\textit{et al.}\cite{27} proposed a scheme of PA+BS, and it can also beat the
SQL of phase sensitivity by a similar amount for SU(1,1) interferometer.
Szigeti \textit{et al.}\cite{28} presented a pumped-up interferometer where
all the input particles participate in the phase measurement. Anderson
\textit{et al.}\cite{29} constructed a truncated SU(1,1) interferometer in
which the second nonlinear interaction is replaced with balanced homodyne
detection. More recently, Du \textit{et al.}\cite{30} reported a SU(2)-in-SU(1,1)
nested interferometer, which combines advantages of SU(1,1) and SU(2) interferometry.
Zuo \textit{et al.}\cite{31} proposed and experimentally demonstrated a compact
quantum interferometer combining squeezing and parametric amplification.

Besides the phase estimation, the angular displacement estimation has been
another topic of interest with a number of potential applications, including rotational control
of microscopic systems \cite{32}, detecting spinning objects \cite{33}
and exploration of effects such as the rotational Doppler shift \cite{34},
and so on \cite{35,36}. The use of light endowed with  orbital angular momentum (OAM) 
can improve the sensitivity of the angular displacement measurement, which amplifies 
an angular displacement $\theta$ to $l\theta$ \cite{37}, where $l$ denotes topological 
charge and can take any integer value. Recently, some interferometer configurations
have been utilized to realize precision measurement of angular displacement.
Jha \textit{et al.}\cite{38} showed that the sensitivity of angular
displacement in MZI can reach $\frac{1}{2l\sqrt{N}}$ and $\frac{1}{2lN}$ by
employing $N$-unentangled photons and $N$-entangled photons, respectively. Liu
\textit{et al.}\cite{39} analyzed the sensitivity of angular displacement based on
an SU(1,1) interferometer. Zhang \textit{et al.}\cite{40} investigated angular
displacement estimation via the scheme of PA+BS. In addition, some estimation protocols using other inputs and detection strategies are studied\cite{42,43,44}. Based on the modified
MZI\cite{31}, we can obtain the higher sensitivity of angular displacement for the case of angular displacement only in one arm. Furthermore, if two angular displacements are introduced to both arms, no researchers have studied this situation from the multiparameter estimation perspective so far.

In this paper, we study the estimation of angular displacements based on a modified MZI. The multiparameter quantum Cram\'{e}r-Rao bound (QCRB) is derived using the method of quantum Fisher information matrix (QFIM), and the bound of angular displacements difference between the two arms is compared with the sensitivity of angular displacement using the intensity detection. For the angular displacement is only in one arm, the QCRB of angular displacement is compared with the sensitivity of the homodyne detection method.

\begin{figure}[tb]
\centering{\includegraphics[scale=0.46,angle=0]{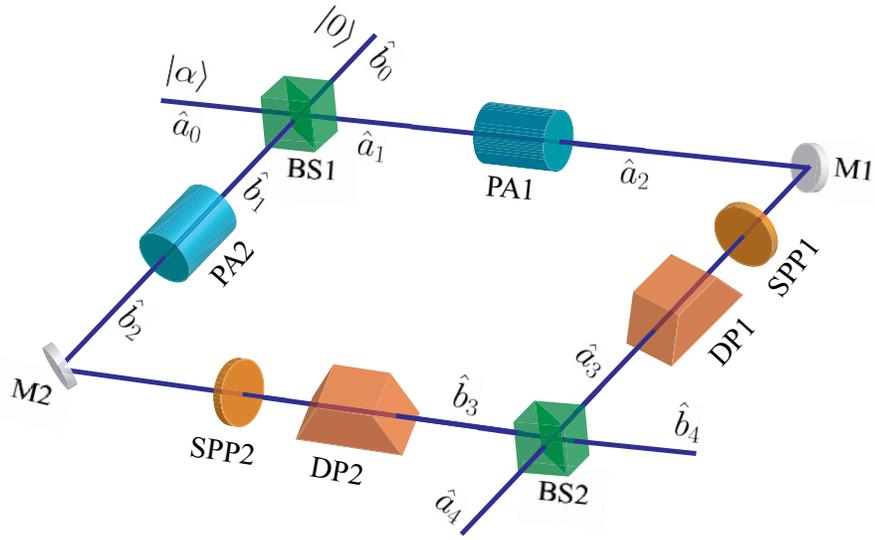}}
\caption{Two sets of optical parametric amplifiers (PAs), spiral phase
plates (SPPs) and Dove prisms (DPs) are placed in two arms of the standard MZI,
respectively. The squeezed states generated from the PAs are directly
used for the probe states. The optical field passing through the SPP and DP will have a
phase shift of $2l\theta$, where $l$ denotes topological charge and $\theta$ is the rotation angle of a Dove prism. $\hat{a}_{i}$ and $\hat{b}_{i}$ ($i=0,1,2,3,4$) denote light beams in
the different processes. $\mathrm{M}$: mirrors; $\mathrm{BS}$: beam splitters.}%
\label{fig1}%
\end{figure}

\section{Model}

Different from the usual MZI, two sets of PAs, spiral phase
plates (SPPs) and Dove prisms (DPs) are placed in
two arms of the MZI, respectively, as shown in Fig.~\ref{fig1}. A
coherent state $|\alpha\rangle$ ($\alpha=|\alpha|$) and a vacuum state are
injected into the interferometer. $\hat{b}_{0}$ is in vacuum state, $\hat
{a}_{i}$ and $\hat{b}_{i}$ ($i=0,1,2,3,4$)\ denote light fields in the
different processes. The SPP is used to introduce the OAM degree of freedom, i.e. the light field passing through the SPP
will carry OAM. The employment of DP transforms the topological charge from
$l$ to $-l$ and imposes a phase shift of $2l\theta$ to the field, where
$\theta$ is the rotation angle of the DP and the parameter to be estimated in
this paper.

The input-output relation of the first beam splitter is given by%
\begin{equation}
\hat{a}_{1}=\sqrt{T}\hat{a}_{0}-\sqrt{R}\hat{b}_{0},\hat{b}_{1}=\sqrt{R}%
\hat{a}_{0}+\sqrt{T}\hat{b}_{0},
\end{equation}
where $R$ and $T$ are the reflectivity and transmissivity of the BS, respectively.
The PAs located in each arm are used for squeezing the shot noise and
amplifying the internal photon number. The relationship between input and
output is%
\begin{equation}
\hat{a}_{2}=\cosh r\hat{a}_{1}+\sinh r\hat{a}_{1}^{\dagger},\hat{b}_{2}=\cosh
r\hat{b}_{1}+\sinh r\hat{b}_{1}^{\dagger},
\end{equation}
where $r$ is the squeezing factor of PAs.
The optical field passing through the SPP and DP is described as%
\begin{equation}
\hat{a}_{3}=\hat{a}_{2}e^{2il\theta_{a}},\hat{b}_{3}=\hat{b}_{2}%
e^{2il\theta_{b}},
\end{equation}
where $\theta_{a}$ and$\ \theta_{b}$ correspond to the rotation angles of the
DP1 and DP2, respectively.
The full\ input-output relation of the scheme is given by
\begin{align}
\hat{a}_{4} &  =(Te^{2il\theta_{a}}+Re^{2il\theta_{b}})(\cosh r\hat{a}%
_{0}+\sinh r\hat{a}_{0}^{\dagger})\nonumber\\
&  +\sqrt{TR}(e^{2il\theta_{b}}-e^{2il\theta_{a}})(\cosh r\hat{b}_{0}+\sinh
r\hat{b}_{0}^{\dagger}),\nonumber\\
\hat{b}_{4} &  =(Te^{2il\theta_{b}}+Re^{2il\theta_{a}})(\cosh r\hat{b}%
_{0}+\sinh r\hat{b}_{0}^{\dagger})\nonumber\\
&  +\sqrt{TR}(e^{2il\theta_{b}}-e^{2il\theta_{a}})(\cosh r\hat{a}_{0}+\sinh
r\hat{a}_{0}^{\dagger}).\label{eq4}%
\end{align}

\section{Angular displacements estimation}

\subsection{Angular displacements in both arms}

In this section, we consider a general situation in which $\theta_{d}$ and
$\theta_{s}$ are the parameters to be estimated, where $\theta_{d}=\theta
_{b}-\theta_{a}$, $\theta_{s}=\theta_{b}+\theta_{a}$. Such a kind of problem
can be dealt with by the multiparameter quantum estimation theory. The bounds
of the angular displacement estimation uncertainties can be obtained by the
quantum Cram\'{e}r-Rao inequality\cite{45,46,41}:%
\begin{equation}
\Sigma(\theta_{1},\theta_{2})\geq\mathcal{F}^{-1}(\theta_{1},\theta_{2}),
\end{equation}
where $\Sigma$ is the covariance matrix for parameters $\theta_{1}$,
$\theta_{2}$, $\mathcal{F}^{-1}$ is the inverse matrix of the QFIM
$\mathcal{F}(\theta_{1},\theta_{2})$ with elements $\mathcal{F}_{ij}$
($i,j=1,2$) given by%
\begin{equation}
\mathcal{F}_{ij}=\mathrm{Tr}\left[  \rho(\theta_{1},\theta_{2})\frac{\hat
{L}_{i}\hat{L}_{j}+\hat{L}_{j}\hat{L}_{i}}{2}\right]  ,\label{eq6}%
\end{equation}
in which $\rho$ is the density matrix of the system and the symmetrized
logarithmic derivatives $\hat{L}_{i}$ defined by%
\begin{equation}
\frac{\partial\rho(\theta_{1},\theta_{2})}{\partial\theta_{i}}=\frac{\rho
\hat{L}_{i}+\hat{L}_{i}\rho}{2}.
\end{equation}
For the case of unitary evolution of a pure initial state, the QFIM can be
caiculated analytically,%
\begin{equation}
\mathcal{F}_{ij}=4\operatorname{Re}(\langle\partial_{i}\psi_{\theta}%
|\partial_{j}\psi_{\theta}\rangle-\langle\partial_{i}\psi_{\theta}%
|\psi_{\theta}\rangle\langle\psi_{\theta}|\partial_{j}\psi_{\theta}%
\rangle),\label{eq2}%
\end{equation}
where $|\psi_{\theta}\rangle$ is the state vector after evolving through DP
and $|\partial_{i}\psi_{\theta}\rangle=\partial|\psi_{\theta}\rangle
/\partial\theta_{i}$.

For the estimation of $\theta_{d}$ and $\theta_{s}$, the QFIM is given by%
\begin{equation}
\mathcal{F}(\theta_{d},\theta_{s})=\left(
\begin{array}
[c]{cc}%
F_{dd} & F_{{ds}}\\
F_{sd} & F_{ss}%
\end{array}
\right)  ,
\end{equation}
where the subscripts $d$ and $s$ denote $\theta_{d}$ and $\theta_{s}.$
Using Eq.~(\ref{eq2}) and Eq.~(\ref{eq4}), the matrix elements take the form%
\begin{align}
F_{dd} &  =4l^{2}[\langle(\hat{n}_{b_{2}}-\hat{n}_{a_{2}})^{2}\rangle
-\langle\hat{n}_{b_{2}}-\hat{n}_{a_{2}}\rangle^{2}]\nonumber\\
&  =4l^{2}(\left\vert \alpha\right\vert ^{2}e^{4r}+\sinh^{2}2r),\nonumber\\
F_{ds} &  =4l^{2}[\langle(\hat{n}_{b_{2}}-\hat{n}_{a_{2}})(\hat{n}_{b_{2}%
}+\hat{n}_{a_{2}})\rangle-\langle\hat{n}_{b_{2}}-\hat{n}_{a_{2}}\rangle
\langle\hat{n}_{b_{2}}+\hat{n}_{a_{2}}\rangle]\nonumber\\
&  =4l^{2}\left\vert \alpha\right\vert ^{2}e^{4r}(1-2T),\nonumber\\
F_{sd} &  =4l^{2}[\langle(\hat{n}_{b_{2}}+\hat{n}_{a_{2}})(\hat{n}_{b_{2}%
}-\hat{n}_{a_{2}})\rangle-\langle\hat{n}_{b_{2}}+\hat{n}_{a_{2}}\rangle
\langle\hat{n}_{b_{2}}-\hat{n}_{a_{2}}\rangle]\nonumber\\
&  =4l^{2}\left\vert \alpha\right\vert ^{2}e^{4r}(1-2T),\nonumber\\
F_{ss} &  =4l^{2}[\langle(\hat{n}_{b_{2}}+\hat{n}_{a_{2}})^{2}\rangle
-\langle\hat{n}_{b_{2}}+\hat{n}_{a_{2}}\rangle^{2}]\nonumber\\
&  =4l^{2}(\left\vert \alpha\right\vert ^{2}e^{4r}+\sinh^{2}2r),
\end{align}
where $\hat{n}_{a_{2}}=\hat{a}_{2}^{\dagger}\hat{a}_{2}$, $\hat{n}_{b_{2}%
}=\hat{b}_{2}^{\dagger}\hat{b}_{2}$.
Then the corresponding bounds are given by%
\begin{align}
\Delta^{2}\theta_{d}  &  \geq\frac{F_{ss}}{F_{dd}F_{ss}-F_{ds}F_{sd}%
},\nonumber\\
\Delta^{2}\theta_{s}  &  \geq\frac{F_{dd}}{F_{dd}F_{ss}-F_{ds}F_{sd}}.
\label{eq3}%
\end{align}
As seen in Eq.~(\ref{eq3}), in general,\ the bound for\ estimating $\theta
_{d}$ depends on the information amount\ of $\theta_{s}$ due to the nonzero
off-diagonal elements in the QFIM. When $R=T=1/2$, $F_{ds}=F_{sd}%
=0$,\ then~Eq.~(\ref{eq3}) is reduced to%
\begin{align}
\Delta^{2}\theta_{d} &  \geq\frac{1}{F_{dd}},\nonumber\\
\Delta^{2}\theta_{s} &  \geq\frac{1}{F_{ss}},\label{eq5}%
\end{align}
which means that one does not need to consider if $\theta_{s}$ is known to find the
ultimate bound of $\theta_{d}$.

The QFI is the intrinsic information in the
quantum state and is not related to the actual measurement procedure. It
characterizes the maximum amount of information that can be extracted from
quantum experiments about an unknown parameter using the best (and ideal)
measurement device. Here, we consider the intensity detection as our
measurement strategy, and compare the bound of $\theta_{d}$ in Eq.~(\ref{eq5})
with the sensitivity using the intensity detection.

The sensitivity is obtained through an error propagation analysis%
\begin{equation}
\Delta\theta=\frac{(\Delta^{2}\hat{O})^{1/2}}{|\partial\langle\hat{O}%
\rangle/\partial\theta|},
\end{equation}
where $\Delta^{2}\hat{O}$ and $|\partial\langle\hat{O}\rangle/\partial
\theta_{d}|$ denote the noise of observable $\hat{O}$ and its rate of change
with respect to $\theta$, respectively. The detected variable $\hat{O}$ can be
phase quadrature or the photon number. Here, we use the difference in
the\ intensities of the two output ports as the detection variable, that is%
\begin{align}
\hat{N} &  =\hat{a}_{4}^{\dagger}\hat{a}_{4}-\hat{b}_{4}^{\dagger}\hat{b}%
_{4}=\mathcal{A}\hat{T}_{1}+\mathcal{B}\hat{T}_{2}+\mathcal{B}^{\ast}\hat
{T}_{2}^{\dagger}-\mathcal{A}\hat{T}_{3},
\end{align}
where%
\begin{align}
\mathcal{A} &  =(T-R)^{2}+4TR\cos(2l\theta_{d}),\mathcal{B}=2\sqrt
{TR}(R-Re^{-2il\theta_{d}}+Te^{2il\theta_{d}}-T),\nonumber\\
\hat{T}_{1} &  =\cosh^{2}r\hat{a}_{0}^{\dagger}\hat{a}_{0}+\cosh r\sinh
r\hat{a}_{0}^{\dagger2}+\cosh r\sinh r\hat{a}_{0}^{2}+\sinh^{2}r\hat{a}%
_{0}\hat{a}_{0}^{\dagger},\nonumber\\
\hat{T}_{2} &  =\cosh^{2}r\hat{a}_{0}^{\dagger}\hat{b}_{0}+\cosh r\sinh
r\hat{a}_{0}^{\dagger}\hat{b}_{0}^{\dagger}+\cosh r\sinh r\hat{a}_{0}\hat
{b}_{0}+\sinh^{2}r\hat{a}_{0}\hat{b}_{0}^{\dagger},\nonumber\\
\hat{T}_{3} &  =\cosh^{2}r\hat{b}_{0}^{\dagger}\hat{b}_{0}+\cosh r\sinh
r\hat{b}_{0}^{\dagger2}+\cosh r\sinh r\hat{b}_{0}^{2}+\sinh^{2}r\hat{b}%
_{0}\hat{b}_{0}^{\dagger}.
\end{align}

Under the\ condition of $R=T=1/2$, the sensitivity of angular displacement can
be written as%
\begin{equation}
\Delta\theta_{d}^\mathrm{ID}=\frac{\sqrt{|\alpha|^{2}[\cos^{2}(2l\theta_{d}%
)e^{4r}+\sin^{2}(2l\theta_{d})]+\sinh^{2}(2r)\cos^{2}(2l\theta_{d})}%
}{2l|\alpha|^{2}e^{2r}|\sin(2l\theta_{d})|}.
\end{equation}
where the superscript $\mathrm{ID}$ denotes the intensity detection.
When $2l\theta_{d}=\pi/2$, the optimal sensitivity is%
\begin{equation}
\Delta\theta_{d_{optimal}}^\mathrm{ID}=\frac{1}{2l|\alpha|e^{2r}}.
\end{equation}
Compared with the bound for\ estimating $\theta_{d}$ in Eq.~(\ref{eq5}), we
find that when the intensity of the input coherent state is strong enough,
$\Delta\theta_{d_{optimal}}^\mathrm{ID}\approx\Delta\theta_{d}$, i.e. the optimal
sensisitivity using the intensity detection can saturate the corresponding QCRB.

\subsection{Angular displacement in one arm}

In the preceding section, we considered a two-parameter estimation problem and
used the QFIM approach to calculate the QCRB of the angular displacements.
Now we assume $R=T=1/2$, $\theta_{a}=0$ and\ calculate the single parameter
QCRB. For the single parameter estimation, the QCRB according to the QFI is
given by \cite{41}%
\begin{equation}
\Delta\theta_\mathrm{QCRB}=\frac{1}{\sqrt{\mathcal{F}}}.\label{eq7}%
\end{equation}
Eq.~(\ref{eq6}) is simplified to%
\begin{equation}
\mathcal{F}=\mathrm{Tr}[\rho(\theta)L^{2}],
\end{equation}
where the corresponding symmetrized logarithmic derivatives $L$ defined by%
\begin{equation}
\frac{\partial\rho(\theta)}{\partial\theta}=\frac{\rho L+L\rho}{2}.
\end{equation}

Under the lossless condition, for a pure state, the QFI is reduced to%
\begin{equation}
\mathcal{F}=4(\langle\partial_{\theta}\psi_{\theta}|\partial_{\theta}\psi_{\theta
}\rangle-|\langle\partial_{\theta}\psi_{\theta}|\psi_{\theta}\rangle|^{2}),
\end{equation}
where $|\psi_{\theta}\rangle$ is the state vector after evolving through DP
and $|\partial_{\theta}\psi_{\theta}\rangle=\partial|\psi_{\theta}%
\rangle/\partial\theta$.
In our case, the QFI is given by%
\begin{equation}
\mathcal{F}=16l^{2}\left[ \langle \hat{n}_{b_{2}}^{2}\rangle -\langle \hat{n}%
_{b_{2}}\rangle ^{2}\right] =8l^{2}(e^{4r}\left\vert \alpha \right\vert
^{2}+\sinh ^{2}2r).
\end{equation}
Using Eq.~(\ref{eq7}), the QCRB is%
\begin{equation}
\Delta\theta_\mathrm{QCRB}=\frac{1}{2\sqrt{2}l\sqrt{e^{4r}\left\vert \alpha
\right\vert ^{2}+\sinh^{2}2r}}.
\end{equation}

Then we analyze the measurement sensitivity of angular displacement $\theta$
($\theta=\theta_{b}$) using the balanced homodyne detection. The measurement
operator is%
\begin{equation}
\hat{Y}_{b_{4}}=-i(\hat{b}_{4}-\hat{b}_{4}^{\dagger}).
\end{equation}
When $R=T=1/2$ and $\theta_{a}=0$, Eq.~(\ref{eq4}) can be written as%
\begin{align}
\hat{a}_{4}& =[(e^{2il\theta }+1)(\cosh r\hat{a}_{0}+\sinh r\hat{a}%
_{0}^{\dagger })+(e^{2il\theta }-1)(\cosh r\hat{b}_{0}+\sinh r\hat{b}%
_{0}^{\dagger })]/2,  \nonumber \\
\hat{b}_{4}& =[(e^{2il\theta }+1)(\cosh r\hat{b}_{0}+\sinh r\hat{b}%
_{0}^{\dagger })+(e^{2il\theta }-1)(\cosh r\hat{a}_{0}+\sinh r\hat{a}%
_{0}^{\dagger })]/2.
\end{align}
The rate of change of $\langle\hat{Y}_{b_{4}}\rangle$ with respect to $\theta$
and the variance of $\hat{Y}_{b_{4}}$ is given by%
\begin{equation}
|\partial_{\theta}\langle\hat{Y}_{b_{4}}\rangle|=2l(\cosh r+\sinh
r)|\alpha\cos(2l\theta)|,
\end{equation}
and%
\begin{equation}
\Delta^{2}\hat{Y}_{b_{4}}=-\sinh r\cosh r[\cos(4l\theta)+1]+\cosh(2r),
\end{equation}
respectively.
Using error propagation formula, the sensitivity is%
\begin{equation}
\Delta\theta^\mathrm{BHD}=\frac{\sqrt{-\sinh r\cosh r[\cos(4l\theta)+1]+\cosh(2r)}%
}{2l(\cosh r+\sinh r)|\alpha\cos(2l\theta)|},
\end{equation}
where the superscript $\mathrm{BHD}$ denotes the balanced homodyne detection.

\begin{figure}[b]
\centering{\includegraphics[scale=0.6,angle=0]{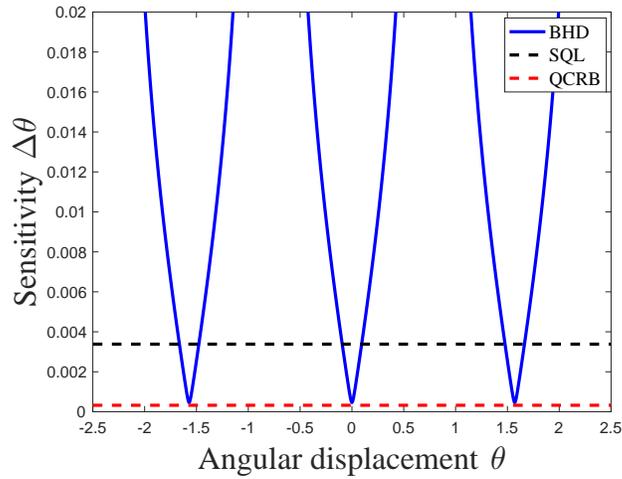}}
\caption{The sensitivity as a function of angular displacement $\protect%
\theta$ in the case of $l=1$, $r=2$ and $|\protect\alpha|=20$.}
\label{fig2}
\end{figure}

Fig.~\ref{fig2} shows the behavior of $\Delta\theta^\mathrm{BHD}$ as a function of $\theta
$. This result shows that the sensitivity of balanced homodyne detection can
surpass the SQL and approach the QCRB. When $2l\theta=k\pi$ ($k$ is an
arbitrary integer), the fluctuation of $\hat{Y}_{b_{4}}$ is reduced to
$e^{-2r}$\ and the optimal sensitivity is obtained%
\begin{equation}
\Delta\theta_{optimal}^\mathrm{BHD}=\frac{\cosh r-\sinh r}{2l(\cosh r+\sinh r)|\alpha|}.
\label{eq1}%
\end{equation}

\begin{figure}[tbp]
\centering{\includegraphics[scale=0.6,angle=0]{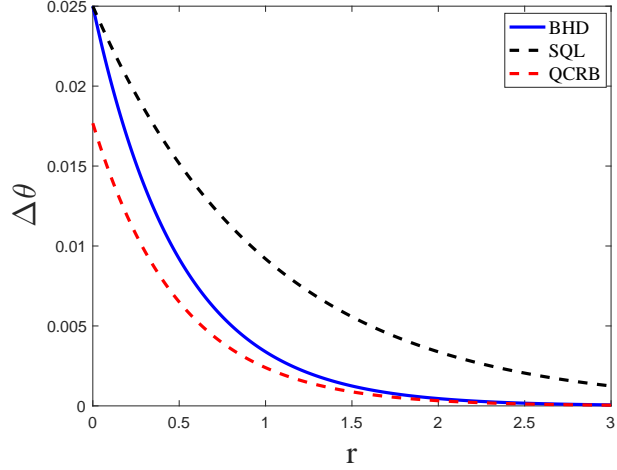}}
\caption{The effect on the phase sensitivity with the increase in the
squeezing parameter $r$. Plotted with $\protect\theta=0$, $l=1$ and $|%
\protect\alpha|=20$.}
\label{fig3}
\end{figure}

Next, we show the effect of the increase in the squeezing parameter $r$ of PAs
on the sensitivity of angular displacement. In Fig.~\ref{fig3}, we see that with increase in $r$, 
the sensitivity of our scheme improves. And the sensitivity always goes below the SQL.
With higher $r$, the sensitivity of our scheme keeps increasing and approaches the QCRB.

Due to the amplification process, the phase-sensing photon number is the total number of photons inside the interferometer, not the input photon number as the traditional MZI, that is%
\begin{equation}
N_{Tot}=(\cosh r+\sinh r)^{2}|\alpha|^{2}+2\sinh^{2}r.
\end{equation}
Eq.(\ref{eq1}) can be written in terms of the internal number of photons, such
that%
\begin{equation}
\Delta\theta_{optimal}=\sqrt{\frac{(\cosh r-\sinh r)^{2}}{4l^{2}%
(N_{Tot}-2\sinh^{2}r)}}\approx\sqrt{\frac{1}{4l^{2}N_{Tot}(\cosh r+\sinh
r)^{2}}}\approx\frac{1}{4l\cosh r}\sqrt{\frac{1}{N_{Tot}}}.
\end{equation}
The angular displacement sensitivity can be enhanced by a factor of $2\cosh r$
compared with the SQL $(1/2l\sqrt{N})$ under the condition of $N_{Tot}\gg\sinh^{2}r$ and $\cosh
r\gg1$. When $\sinh r\gg1$ and $\sinh^{2}r=N_{Tot}/4$, the optimal sensitivity
of the scheme can reach the Heisenberg limit, approximately as $1/2lN$.

\begin{figure}[tbp]
\centering{\includegraphics[scale=0.6,angle=0]{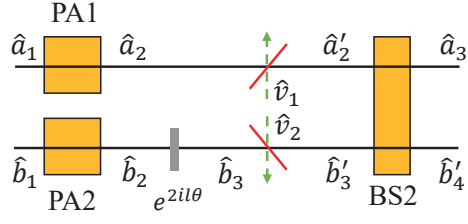}}
\caption{A lossy interferometer model, the photon losses are modeled by
adding fictitious beam splitters.}
\label{fig4}
\end{figure}

Additionally, we investigate the effect of photon losses on sensitivity.
Losses can be modeled by adding fictitious beam splitters, as shown in
Fig.~\ref{fig4}. Considering two arms of the interferometer have the same
transmission rates $\eta$, the optical fields $\hat{a}_{2}$ and $\hat{b}_{3}$
suffering from photon losses are given by%
\begin{equation}
\hat{a}_{2}^{^{\prime}}=\sqrt{\eta}\hat{a}_{2}+\sqrt{1-\eta}\hat{v}_{1}%
,\hat{b}_{3}^{^{\prime}}=\sqrt{\eta}\hat{b}_{3}+\sqrt{1-\eta}\hat{v}_{2},
\end{equation}
where $\hat{v}_{1}$ and $\hat{v}_{2}$ represent vacuum.

Then, the full\ input-output relation is described as%
\begin{align}
\hat{b}_{4}^{^{\prime }}& =\sqrt{\eta }[(e^{2il\theta }+1)(\cosh r\hat{b}%
_{0}+\sinh r\hat{b}_{0}^{\dagger })+(e^{2il\theta }-1)(\cosh r\hat{a}%
_{0}+\sinh r\hat{a}_{0}^{\dagger })]/2  \nonumber \\
& +\sqrt{1-\eta }(\hat{v}_{2}-\hat{v}_{1})/\sqrt{2}.
\end{align}
After taking into account the photon losses, the sensitivity is given by%
\begin{equation}
\Delta\theta^{^{\prime}}=\frac{\sqrt{-\eta\cosh r\sinh r[\cos(4l\theta
)+1]+\eta\cosh(2r)+1-\eta}}{2\sqrt{\eta}l(\cosh r+\sinh r)|\alpha\cos
(2l\theta)|}.
\end{equation}

\begin{figure}[tbp]
\centering{\includegraphics[scale=0.6,angle=0]{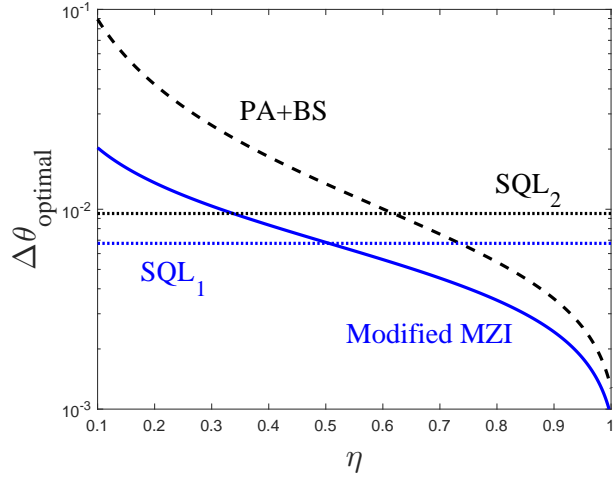}} \caption{The optimal
sensitivity versus photon losses coefficient $\eta$, where $l=1$, $r=2$ and
$|\alpha|=10$. The solid line and dashed line represent the sensitivity of our scheme and PA+BS scheme, respectively. SQL$_{1}$ and SQL$_{2}$ correspond to
respective standard quantum limit.}%
\label{fig5}%
\end{figure}

In Ref.\cite{40}, the author calculated the sensitivity of homodyne detection
with photon losses in PA+BS scheme, it can resist $38\%$ photon losses in the case
of $r=2$, $l=1$, and $|\alpha|=10$. Here, a comparison between our scheme and
PA+BS scheme is provided under the same condition. As shown in Fig.~\ref{fig5},
we plot the optimal sensitivity as a function of photon losses coefficient.
The solid line and dashed line represent the sensitivity of our scheme and PA+BS scheme, respectively. SQL$_{1}$ and SQL$_{2}$ correspond to respective standard quantum limit. Our scheme is robust against photon losses, it can tolerate approximately photon losses of $50\%$. Besides, it should be noted that the sensitivity curve is always below PA+BS scheme, which is due to the amplification of the internal photon number.

\begin{table}[htbp]
    \centering
    \caption{The sensitivities of angular displacement for different interferometer configurations with coherent state $\otimes$ vacuum state input and balanced homodyne detection}
    \begin{tabular}{cc}
         \toprule
         Interferometer configurations & Sensitivities of angular displacement  \\
         \midrule
         standard MZI & $1/2lN_{\alpha}^{1/2}$  \\
         PA+PA scheme & $1/2l\mathcal{N}N_{\alpha}^{1/2}$  \\
         PA+BS scheme & $1/[2\sqrt{2}l(N_{r}/2+\mathcal{N}/2+1)N_{\alpha}^{1/2}]$\cite{40}  \\
         modified MZI & $1/[2l(N_{r}+\mathcal{N}+1)N_{\alpha}^{1/2}]$  \\
         \bottomrule

    \end{tabular}
\label{Tab1}
\end{table}
In Table \ref{Tab1}, we summarize the sensitivities of angular displacement for different interferometer configurations with coherent state $\otimes$ vacuum state input and balanced homodyne detection. The sensitivity of the PA+PA scheme is higher than that of the standard MZI by a factor of $\mathcal{N}$, where $\mathcal{N}=[(N_{r}+2)N_{r}]^{1/2}$, and $N_{\alpha}=|\alpha|^{2}$ is the mean photon number of input coherent state, $N_{r}=2\sinh^{2}r$ is the spontaneous photon number emitted from the PA. When $\sinh r\gg1$, the sensitivity of PA+BS scheme is greater than the PA+PA scheme by a factor of $\sqrt{2}$\cite{40}. The sensitivity of angular displacement in our scheme is further improved by a factor of $2$ compared to the PA+PA scheme, and the enhancement factor results from both effects of amplified internal photon number and squeezed noise.

\section{Conclusion}

In conclusion, we have investigated the estimation of angular displacements based on
a modified MZI. When the unknown angular displacements are in both arms, the sensitivity using  intensity detection can saturate the QCRB of angular displacements difference obtained by using the method of QFIM. For the angular displacement is only in one arm, the sensitivity using the method of homodune detection can be enhanced by a factor of $2\cosh r$ compared with the SQL\ and approach the QCRB. Additionally, our sheme can tolerate approximately $50\%$ photon losses. We summarize the sensitivities of angular displacement for different interferometer configurations, the sensitivity of angular displacement in our scheme is improved due to the reduction of shot noise and amplification of photon number inside the interferometer. It will have potential applications in quantum sensing and precision measurements.

\section*{Funding}
This work is supported by the National Natural Science Foundation of China (Grant Nos. 11974111, 11874152, 11604069, 91536114, 11654005, and 11234003, 11474095), the Fundamental Research Funds for the Central Universities, the Science Foundation of Shanghai, China (Grant No. 17ZR1442800), and the National Key Research and Development Program of China (Grant No. 2016YFA0302001).

\section*{Disclosures}
The authors declare no conflicts of interest.


\end{document}